\documentclass[pdflatex,sn-basic]{sn-jnl}% Basic Springer Nature Reference Style/Chemistry Reference Style

\begin{document}

\title{The origin of free-floating planets}

\author{\fnm{Núria} \sur{Miret-Roig}}\email{nuria.miret.roig@univie.ac.at}

\affil{University of Vienna, Department of Astrophysics, Türkenschanzstraße 17, 1180 Wien, Austria}

\abstract{Free-floating planets (FFPs) are the lightest products of star formation and they carry important information on the initial conditions of the environment in which they were formed. They were first discovered in the 2000s but still few of them have been identified and confirmed due to observational challenges. This is a review of the last advances in the detection of these objects and the understanding of their origin. Several studies indicate that the observed fraction of FFPs outnumbers the prediction of turbulent fragmentation and suggest that many were formed in planetary systems that were later abandoned. The JWST will certainly constitute a new step further in the detection and characterisation of FFPs. To interpret these new observations, precise ages for the nearby star-forming regions in which they were formed will be necessary.}


\maketitle

\footnote{This is a peer-reviewed publication complementing the plenary lecture \textit{MERAC Prize in Observational Astrophysics--Discovery of a large sample of free-floating planets} (No. 856) of the European Astronomical Society Meeting 2022, 27 June -- 1 July 2022, Valencia (Spain).}
 
\section{Introduction}\label{sec1}

Stars, brown dwarfs and planets form in groups which share the same properties (e.g. kinematics, chemical composition) of their parent molecular clouds. The identification of co-eval stars is a very important topic for many astrophysical processes such as planet formation, disc evolution and Galactic dynamics. Now, thanks to the \textit{Gaia} satellite \citep{GaiaColPrusti+16, GaiaColVallenari+2022}, we have excellent astrometry for more than two billion sources which led to the discovery of many new open clusters  (see e.g. \citealt{Castro-Ginard+2022, Cantat-Gaudin+2020}, and references therein), and a better characterisation of young associations and star-forming regions (see e.g. \citealt{Prisinzano+2022, Kerr+2021, Gagne2018a}, and references therein). However, the least massive objects in these complexes, brown dwarfs and free-floating planets, still escape from the \textit{Gaia} detection limit in most cases. They are particularly interesting for studies of the mass function and they carry important information on the initial conditions of star and planet formation. 

The existence of brown dwarfs was predicted in the 60s when \citet{Kumar+1963} and \citet{Hayashi+1963} realised that there was a mass threshold below which a self-gravitating object cannot stably fuse hydrogen. The reason is that the electron degeneracy pressure stops the gravitational collapse before the interior temperatures are high enough to fuse hydrogen. These objects were initially named \textit{black dwarfs} and it was later that the term \textit{brown dwarf} was introduced by \citet{Tarter1975}. The first observational confirmation that such objects indeed exist came with the discovery of Gliese~229B, a brown dwarf orbiting an M dwarf star \citep{Nakajima+1995, Oppenheimer+1995} and Teide~1, an isolated member of the Pleiades cluster \citep{Rebolo+1995, Rebolo+1996}. The first spectroscopic binary brown dwarf, PPl~15 \citep{Basri+1996, Basri+1999}, was discovered soon after. Since then, thousands of brown dwarfs have been detected composing a large sample to study and characterise these objects \citep{Bejar+1999, Zapatero-Osorio+2002, Caballero+2007, Marsh+2010b, Luhman+2013, Smart+2017, Luhman+2018, Kirkpatrick+2019, Kirkpatrick+2021, GaiaColSmart+2021}. 

Despite the enormous progress achieved in the last decades in the understanding of the physics of brown dwarfs, there is still a large number of open questions. Among them, is an accurate definition of the term brown dwarf. Nowadays, the boundary between stars and brown dwarfs is established by the hydrogen-burning limit which, according to the models, occurs around $75$~M$_\textup{J}$ \citep{Burrows+1993}. Objects above this threshold are massive enough to maintain hydrogen nuclear reactions and become stars. On the contrary, brown dwarfs never reach a temperature high enough to initiate the hydrogen burning in their interiors and spend their whole existence slowly contracting and cooling. Nonetheless, they can temporarily have nuclear reactions such as deuterium burn. However, this definition based on the mass has caveats (see e.g. \citealt{Chabrier+1997, Forbes+2019}).

The boundary between brown dwarfs and planets is even more controversial. The Working Group on Extrasolar Planets of the International Astronomical Union (IAU) established the deuterium burning limit ($\sim13$~M$_\textup{J}$) as a mass threshold to distinguish brown dwarfs from lighter isolated objects \citep{Boss+2007}. There is still debate on how to name objects lighter than $13$~M$_\textup{J}$ which are unbound to a more massive object (star or brown dwarf). Some of the names used in the literature are \textit{sub-brown dwarfs}, \textit{isolated planetary-mass objects}, \textit{rogue planets}, \textit{free-floating planets} (FFPs) and, through this article, I use the latter. Some authors have argued that the deuterium burning limit has little impact on the evolution of the object and suggested another division based on the formation mechanism \citep{Chabrier05, Chabrier+2014, Spiegel+2011}. In this perspective, brown dwarfs are substellar objects that form from a turbulent gravitational collapse like low-mass stars. This definition imposes a different low-mass limit known as the opacity limit ($\sim3$~M$_\textup{J}$). In contrast, planets form in circumstellar discs and orbit a more massive object. This second criterion to classify brown dwarfs and planets implies a mass overlap between the two categories and is difficult (if possible) to apply when the formation mechanism is unknown. Some free-floating planets might have formed around a star and have been dynamically ejected \citep{Raymond+2010, Parker+2012, vanElteren2019}. Lacking a precise term to describe the uncertain nature of unbound objects less massive than 13~~M$_\textup{J}$, I follow the guidelines of the IAU and consider brown dwarfs all the objects with masses in the range $13-75$~M$_\textup{J}$ and FFPs the objects with masses $<13$~M$_\textup{J}$, regardless of their formation mechanism.

The first FFPs were detected at the turn of the century in nearby star-forming regions \citep{Lucas+2000, Lucas+2001, Zapatero-Osorio+2000, Oasa+1999, Luhman+2004}. The faintness of these objects makes them extremely difficult to detect, only being possible in very close and young regions (when they are still relatively warm and bright). The most secure methodology to confirm the detection of FFPs is spectroscopy however, these ultra-faint objects require long integration times even on the largest telescopes on Earth. On the lack of spectroscopic observations, precise proper motions can be used together with multi-band photometry to identify co-moving objects candidates of FFPs. Since the first discovery of FFPs, others have been detected in nearby young associations \citep{Lucas+2006, Lodieu+2007, Lodieu+2018, Lodieu+2021, Weights+2009, Scholz+2009, Scholz+2012a, Marsh+2010a, Muzic+2012, Pena-Ramirez+2012, Liu+2013, Faherty+2013, Ingraham+2014, Kellogg+2015, Schneider+2016, Luhman+2016, Gagne+2017, Best+2017, Esplin+2017, Esplin+2019, Zapatero-Osorio+2017}, the solar neighbourhood \citep{Kirkpatrick+2019, Kirkpatrick+2021} and in gravitational microlensing surveys of the Galactic field \citep{Sumi+2011, Mroz+2017, Mroz+2020, Ryu+2021, McDonald+2021}. I refer to \citet{Caballero+2018} for a more extended review of the discovery and properties of substellar objects.  

This article aims at reviewing the latest advances in the detection of the lightest products of star formation and the understanding of their origin. It is particularly focused on my PhD results, presented at the EAS 2022. In Sect.~\ref{sec:formation-mechanisms}, I describe the different scenarios proposed to explain the formation of substellar objects and how the predictions of these theories match with observations. In Sect.~\ref{sec:FFPs-USC}, I present the recent discovery of the largest family of FFPs in Upper Scorpius and Ophiuchus. I discuss the formation mechanisms of this new sample of FFPs by comparing the observed mass function with simulations. The significant uncertainty on the age of this region is one of the main limitations in obtaining the observed mass function. To address this difficulty, in Sect.~\ref{sec:ages}, I discuss the best methodology to determine the age of a young stellar association. In Sect.~\ref{sec:conclusions}, I present the conclusions and future perspectives.

\section{Scenarios of their origin}\label{sec:formation-mechanisms}

There is a puzzle that accompanies the understanding of brown dwarf and FFP formation: they are numerous (almost as much as stars) but have masses two orders of magnitude smaller than the average Jeans mass in star-forming clouds. To obtain lower Jeans masses, the densities of prestellar cores, parents of brown dwarfs, must be high. Alternatively, the accretion has to stop before the prestellar core becomes a low-mass star. Several mechanisms have been proposed in the literature to explain the formation of low-mass stars and brown dwarfs, which are reviewed by \citet{Parker+2020} and \citet{Whitworth18}. The feasibility and contribution of each of them to the final population of brown dwarfs and FFPs are still under debate.

\subsection{Star-like formation -- Turbulent fragmentation}

Turbulent fragmentation is the driving mechanism to form stars. In this scenario, a prestellar core forms when the collision between turbulent flows creates a condensation unstable under the Jeans criterion \citep{Padoan+2002, Padoan+2004, Hennebelle+2008, Hopkins+2012, Padoan+2020}. Some evidence in favour of this scenario is that the disc \citep{Luhman+2005, Monin+2010} and binary \citep{Burgasser+2003, Bouy+2003, Bouy+2006a, Bouy+2006b, Fontanive+2018} properties of substellar objects resemble those of low-mass stars. Additionally, several studies have detected candidates of prestellar cores with a mass below the hydrogen-burning limit and proto-brown dwarfs \citep{Andre+2012, Palau+2012, Palau+2014, Lee+2013, Gregorio-Monsalvo+2016, Riaz+2016, Huelamo+2017, Santamaria-Miranda+2020}.

\subsection{Planet-like formation -- Ejection from planetary system}

Low-mass brown dwarfs and FFPs can form in a circumstellar disc around a star or a massive brown dwarf. This can happen in two different ways.

\textbf{Disc fragmentation}. This mechanism occurs when the circumstellar disc surrounding the primary body fragments, becomes unstable and collapses \citep{Boss1998, Bate+2002}. These fragments continue to accrete material while they may interact with the primary body to which they are bound and with other fragments in the same disc. Eventually, these interactions may end with the ejection of one of these cores, usually the least massive. If the ejected body has enough material to resume accretion and sustain hydrogen fusion it becomes a low-mass star. Otherwise, it becomes a brown dwarf or a FFP, depending on the final mass. Direct imaging observations of massive planets on wide orbits favour a disc fragmentation formation scenario rather than core accretion formation in the inner disc plus outwards scattering \citep{Bailey+2014, Bohn+2020, Bohn+2021, Zhang+2021, Janson+2021}.

\textbf{Solid and gas accretion}. In this scenario, planets are formed in protoplanetary discs around young stars by accretion of solids and gas \citep{Pollack+1996}. Dynamical instabilities in the system due to interactions with other bodies of the same system or due to the close passage of an external star may end with the ejection of one of the planets \citep{Rasio1996, Weidenschilling+1996, Veras+2012}.

\subsection{Halted accretion -- Embryos ejection}

Dynamical interactions among cores which are competing to accrete material from the same parent cloud may end with the ejection of the smallest bodies. If the accretion process stops before the cores are massive enough to begin the hydrogen burning, they become a brown dwarf \citep{Reipurth+2001}. \citet{Bate2009} showed that as a result of the ejection process, the majority of discs are truncated, the fraction of binaries decreases and the velocity dispersion increases. Some studies found hints of truncated discs, suggesting that this mechanism might be important in certain environments \citep{Testi+2016}.

\subsection{Halted accretion -- Photo-erosion}

In this scenario, brown dwarfs form in the vicinity of an O-type star where the radiation is strong enough to ionize and evaporate part of the outer layers of the core. At the same time, it adds pressure to the core so that the central part collapses to form a compact body \citep{Whitworth+2004}. The observation of a proplyd in Orion supports this scenario \citep{Bouy+2009, Hodapp+2009}. However, this mechanism can only explain the presence of brown dwarfs in the vicinity of O-type stars which are rare and thus, cannot be the dominant channel. \\[0.4 cm]

\subsection*{Possibly a combination of several mechanisms}

All these mechanisms are likely to be able to form brown dwarfs and probably FFPs. However, it is still unclear which mechanism dominates the formation of substellar objects. Does this depend on the mass range? Or the environment? To address these questions it is fundamental to compare observations of star-forming regions with numerical simulations. One of the most useful parameters to compare observations and simulations is the mass function since different formation mechanisms predict a different proportion of FFPs. Several studies have suggested that substellar objects are a common output of star formation (see e.g. \citealt{Offner+14}, and references therein). However, identifying and confirming FFPs is operationally challenging and it has been possible only in a few cases. To establish the fraction of FFPs to stars, large samples with robust uncertainties and low contamination rates are needed to provide a statistically significant answer. Recently, some works have measured the mass function of the field population reaching planetary mass objects \citep{Kirkpatrick+2019, Kirkpatrick+2021, Bardalez-Gagliuffi+2019, Chabrier+2023}. However, these mass functions are the combination of different star formation events which happened over several Gyr and thus, might differ from the initial mass function reported by simulations of star formation. For that, in the next section, I describe the recent discovery of the largest population of co-eval FFPs to date and discuss how this can help us to learn about star and planet formation.

\section{The largest family of FFPs to date}\label{sec:FFPs-USC}

\subsection{Astro-photometric identification}

The COSMIC DANCe project\footnote{\url{http://www.project-dance.com/}} (DANCe standing for Dynamical Analysis of Nearby ClustErs) started as a survey to map nearby ($<500$~pc), young ($<500$~Myr) associations and open clusters \citep{Bouy+13}. With wide-field, deep ground-based images in the optical and in the infrared, we can detect objects several orders of magnitude fainter than with {\it Gaia}, down to a few Jupiter masses. This project has collected and analysed several thousands of images (including those in public archives) from different instruments and for a single region. Thanks to the large baseline of the observations (10--20 yrs, depending on the region), the typical precision in proper motions is of $\lesssim1$~mas~yr$^{-1}$. The DANCE catalogues include proper motions and multi-filter photometry for millions of sources which can be used to search for the few thousand young stars. This complex problem was addressed by \citet{Sarro+14}, using an expectation-maximisation algorithm to iteratively look for members with similar proper motions that follow the same empirical isochrone to an initial list of candidate members. This algorithm successfully led to the identification of many new substellar objects in nearby star-forming regions \citep{Bouy+15, Olivares+19, Miret-Roig+19, Galli+2020a, Galli+2020b, Galli+2021a, Galli+2021b}. At the same time, \citet{Olivares+2018b, Olivares+2021, Olivares+2022} developed a Bayesian Hierarchical model to provide a new framework to search for members of open clusters and star-forming regions with a better treatment of the interstellar extinction.

The most recent search for substellar objects in the DANCe project is in the area of Upper Scorpius and Ophiuchus \citep{Miret-Roig+2022a}. This is one of the youngest (3--10~Myr) and closest ($\sim140$~pc) star-forming regions to the Sun and is part of the Scorpius-Centaurus complex. Combining more than 80\,000 wide-field images from 18 different instruments (including ESO~VISTA/VIRCAM, ESO~VST/OmegaCAM, CTIO~Blanco/DECam, KPNO~Mayall/NEWFIRM, CFHT/MegaCam, CFHT/WIRCam, UKIRT/WFCAM, Subaru/HSC among others) we were able to map a region of around 170~deg$^2$. These images include observations led by our team and images in public archives taken over the past 20 years. The processing of these images led to a catalogue of proper motions and $grizYJHKs$ photometry for 28\,062\,542~sources, including objects 5~mag fainter than the \textit{Gaia} detection limit. This catalogue was analysed with the expectation-maximisation algorithm developed by \citet{Sarro+14} and \citet{Olivares+19} letting to the identification of 3\,455 co-moving young objects. From these, between 70 and 170 are FFPs and between 380 and 860 are brown dwarfs, depending on the age assumed (between 3 and 10~Myr).

\subsection{Spectroscopic confirmation}

FFPs do not have thermonuclear reactions to sustain high temperatures for millions of years. Instead, they are doomed to cool down and fade away eternally. This nature poses important observational challenges to their detection. FFPs are only visible when they are close and young (i.e. relatively warm and bright). Additionally, there is a degeneracy between young FFPs and more evolved brown dwarfs. To break this degeneracy, it is fundamental to know the age of the FFP or to carry spectroscopic observations that can detect youth indicators on the spectra.

\citet{Bouy+2022} observed 17 of the FPPs identified by \citet{Miret-Roig+2022a} with low-resolution near-infrared spectra obtained with SWIMS at the 8m telescope Subaru and EMIR at the 10m telescope GTC. They were able to confirm 16 of the 17 observed targets as young ultracool objects based on several spectroscopic criteria. Specifically, they confirmed the FFPs by i) comparing the spectra to young and field (old) M and L-dwarf standards \citep{Burgasser+2017}, ii) searching for youth evidence in the spectra such as the slope of the continuum between the $J$ and $Ks$ band, iii) the sharp $H$-band continuum ($H_{cont}$ index, \citealt{Allers+2013}), and iv) the TLI-g gravity-sensitive indices \citep{Almendros-Abad+2022}. These analyses show that the objects have spectral types between L0 and L6 and masses in the range 0.004–0.013 M$_\odot$, according to evolutionary models. This confirms that the contamination of the FPPs sample of \citet{Miret-Roig+2022a} is very low, of $<6\%$.

\subsection{Excess of FFPs compared to turbulent fragmentation simulations}\label{sec:origin}

\begin{figure}
    \includegraphics[width = \columnwidth]{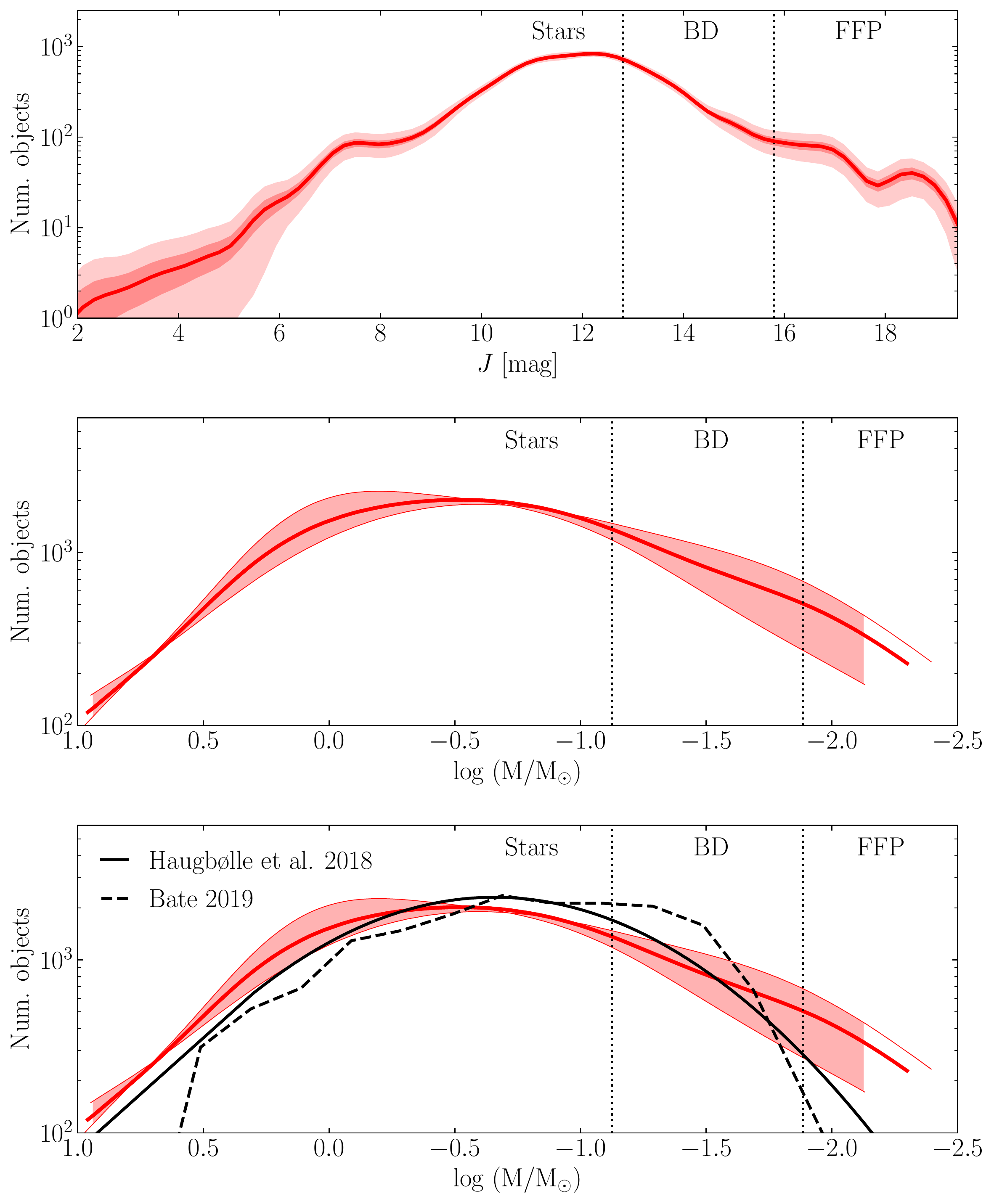}
    \caption{$J$ apparent magnitude distribution and mass functions of the members of Upper Scorpius and Ophiuchus. Top: apparent magnitude distribution. Middle: observational mass function. Bottom: observational mass function (the mass functions from simulations are overlaid). The shaded regions indicate the 1 and 3$\sigma$ uncertainties from a bootstrap (top) and the dispersion due to the age (3--10 Myr, middle and bottom). All of the mass functions are normalised in the mass range 0.004--10~M$_\odot$. The hydrogen (75~M$_\textup{Jup}$) and deuterium (13~M$_\textup{Jup}$) burning limits, which indicate the boundary between stars and brown dwarfs (BD) and BD and FFPs respectively, are indicated by the vertical dotted lines according to the BHAC15 evolutionary models \citep{Baraffe+15} and assuming an age of 5 Myr. Figure from \citet{Miret-Roig+2022a}.}
    \label{fig:mag-mass-func}
\end{figure}

This family of FFPs constitutes the largest and most complete census of co-eval FFPs to date and allows us to investigate their statistical properties. The mass function is an important parameter to compare observations and simulations of star formation, however, both present important challenges. One of the critical points in obtaining the mass function from observations is converting the observed magnitudes to masses. This requires using stellar evolutionary models, which are age-dependent and degenerate for young and substellar objects. At the same time, resolving planetary masses in simulations of star-forming regions is still at the limit of the current computational capabilities. 

Comparing the observational mass function in Upper Scorpius and Ophiuchus with the mass function from different sets of simulations \citep{Bate+2019, Haugbolle+2018}, we found an excess of observed FFPs by a factor of up to seven(see Figure~\ref{fig:mag-mass-func} extracted from \citealt{Miret-Roig+2022a}). These numerical simulations mainly form FFPs similar to stars, by turbulent fragmentation. This excess of observed FFPs compared to simulations indicates that likely other mechanisms also play an important role in FFP formation. In particular, statistics of giant exoplanets and models of planet ejection suggest that a significant fraction of the FFPs observed could have formed by this mechanism. Besides, the dip at planetary masses observed in the model-independent $J$~magnitude distribution could favour this scenario since it could represent the frontier between different formation mechanisms. This dip has also been observed in the Galactic population in the solar neighbourhood \citep{Bardalez-Gagliuffi+2019, GaiaColSmart+2021} but the origin is still unknown. Alternatively, the different fractions of FFPs in observations and simulations could hint that simulations are incomplete at this low-mass regime. The lightest planets detected, of few Jupiter masses, are at the limit of the current resolution of simulations. Constraining the fraction of FFPs that are formed by different mechanisms still requires more effort both from simulations and observations.

\section{The need of accurate ages}\label{sec:ages}

Ages are crucial for understanding most astrophysical processes, for instance, to establish the timescales of planet formation, migration, instabilities and even ejection. In addition, stellar ages are fundamental for determining precise masses of brown dwarfs and FFPs. Lacking precise ages, there is a strong degeneracy between the brightness and the mass of substellar objects. 
The age of Upper Scorpius has been debated for years. Some studies find an age around 5~Myr \citep{Preibisch+2002, David+2019, Miret-Roig+2022b} while others find older values around 10~Myr \citep{Pecaut+2012, Rizzuto+2016, Feiden+2016}. This age difference can represent uncertainties of about 30--40\% on individual masses.
Recently, several studies analysed the 3D spatial distribution and kinematics of the region encompassed by Upper Scorpius and Ophiuchus, identifying multiple sub-populations with an age gradient \citep{Miret-Roig+2022b, Kerr+2021, Squicciarini+2021, Ratzenbock+2022}. This could explain at least part of the discrepancies found by previous authors. Nevertheless, different age tagging techniques can also be responsible for discrepancies up to 50\% \citep[e.g.][]{Barrado2016}. Here, I briefly introduce three of the most used age techniques to determine the stellar age of young stars but there are many other strategies.

\subsection{Isochrone fitting}

One of the most common techniques to determine stellar ages is isochrone fitting. A great advantage of this technique is that it only requires the photometry of a group of stars, which is easy to achieve from an observational point of view. On the contrary, an important caveat is that it relies on evolutionary models which depend on the physics included. They are especially uncertain for very young stars and for very low-mass objects \citep{Baraffe+2002}, where few observational samples are ready to constrain the models. Besides, there are other observational challenges related to dating the ages of young stars with isochrones. Young stars are photometrically variable (due to activity and accretion processes) and are usually affected by interstellar extinction. All this makes difficult the age determination of young stars by isochrone fitting \citep{Jeffries+2014, Jeffries+2021, Binks+2022}. 

\subsection{Lithium depletion boundary}

The Lithium (Li) element is destroyed at temperatures of about $3\cdot10^6$~K. Stellar cores reach this temperature at their interiors at different ages depending on their mass \citep{Bildsten+1997, Ushomirsky+1998}. The lowest mass star which still contains Li provides a good age estimate of the age of a group of stars \citep{Basri+1996, Stauffer+1998, Barrado+2004, Manzi+08, Binks+14}. The main limitation of this technique is from the observational side since Li abundance requires high-resolution spectroscopic observations which are time-consuming for low-mass stars.

\subsection{Dynamical traceback ages}

Dynamical traceback ages constitute a complementary alternative since they are independent of stellar evolutionary models. The 3D positions and velocities of a group of stars and a Galactic 3D potential are sufficient to trace the orbits of individual stars back in time. Before \textit{Gaia} data, the main limitations of this technique were the observational uncertainties which increase with integration time \citep[see e.g.][]{Asiain99, Ortega02, Ducourant14}. The successive \textit{Gaia} Data Releases have provided every time more precise parallaxes and proper motions however, a homogeneous sample of radial velocities of the same precision is still missing. The recent \textit{Gaia} DR3 \citep{GaiaColVallenari+2022} contains the largest sample of homogeneous radial velocities to date but the uncertainties continue to be larger than what we can obtain from the ground with large surveys such as APOGEE \citep{Majewski+2017} or small devoted programs. This improvement in the precision of 6D observables also helps to identify clean samples of co-moving stars and reject possible kinematic contaminants or known spectroscopic binary stars which hinder the traceback analysis. Currently, dynamical traceback ages offer a good opportunity to investigate the ages of young stellar groups of stars \citep{Miret-Roig+18, Miret-Roig+2020b, Miret-Roig+2022b, Kerr+2022a, Kerr+2022b}.

\section{Conclusions \& perspectives}\label{sec:conclusions}

FFPs are a common product of star and planet formation. More than twenty years after their discovery they are still greatly unknown due to strong observational challenges. Up to now, the community has identified a few hundred candidates but only tens have been confirmed with spectroscopy. I have reviewed the last advances in the detection of FFPs and, in particular, the recent identification of 70--170 FFPs in Upper Scorpius and Ophiuchus, 16 of which have been confirmed with spectroscopy, measuring a very low contamination rate of $<6\%$ on the initial sample \citep{Miret-Roig+2022a, Bouy+2022}. This sample constitutes the largest population of young co-eval FFPs to date and is an excellent benchmark to investigate their formation mechanisms. The fraction of FPPs measured in different star-forming regions outnumbers the predictions of a log-normal mass function, suggesting that not all FFPs form as low-mass stars and brown dwarfs but a significant fraction may have been ejected from planetary systems \citep{Scholz+2012a, Pena-Ramirez+2012, Miret-Roig+2022a}. To constrain the fraction of FFPs which have formed by different mechanisms, more efforts from the simulation and observational sides are needed. Simulations that resolve planet formation in discs are necessary to account for a significant fraction of the final FFP simulation. At the same time, we need more and larger samples of observed FFPs to improve the statistical significance of the studies. This should be soon possible thanks to devoted programs of the JWST \citep{Scholz+2022}.

In addition to the mass function, other properties such as the disc and binary fraction can help us to compare observations and simulations. Measuring these properties in brown dwarfs and FFPs is basic to test whether they have similar properties as low-mass stars, indicating a similar formation mechanism or they are very different, suggesting that discs and companions may be perturbed as a result of an ejection process. In this regard, modern facilities such as ALMA, NOEMA and the JWST should allow us to put observational constraints on the disc and binary fraction. Additionally, the JWST will allow us to detect many new FFPs and study for the first time the atmospheres of FFPs and to compare them to giant exoplanets. These observations will certainly help us unravel their origin. 

In parallel to the advances in the detection of FFPs, it is important to determine precise ages for the star-forming regions in which they were born. This is necessary to estimate the masses of these objects, which are otherwise highly degenerated, and to establish the timescale for their formation and ejection process.
\\

\textbf{Acknowledgements}  I thank Nuria Huélamo and Hervé Bouy for their comments on the manuscript. I thank the referee for the useful suggestions that helped to improve the manuscript.

\bibliography{Bibliography}% common bib file

\end{document}